\documentclass[10pt]{article}

\usepackage[a4paper, margin=1in]{geometry}

\usepackage{amsmath, amssymb}

\usepackage{url}
\usepackage{graphicx}
\usepackage{xcolor}
\usepackage[normalem]{ulem}

\usepackage{authblk}

\usepackage[hidelinks]{hyperref}

\usepackage{listings}
\lstdefinelanguage{C}{
	keywords={break, case, catch, continue, debugger, default, delete, do, else, finally, for, function, if, in, instanceof, new, return, switch, throw, try, typeof, var, void, while, with, let, const, class, extends, super, export, import},
	morekeywords={async, await, yield},
	sensitive=true,
	morecomment=[l]{//},
	morecomment=[s]{/*}{*/},
	morestring=[b]",
	morestring=[b]',
	morestring=[b]`
}

\title{The Bloch Equation Generator -- SimuFísica}

\author[1]{M. P. M. de Souza}
\author[1]{G. H. H. Pavão}
\author[2]{A. A. C. de Almeida}
\author[2]{S. S. Vianna}

\affil[1]{Departamento de Física, Universidade Federal de Rondônia, 76900-726, Ji-Paraná, Rondônia, Brazil}
\affil[2]{Departamento de Física, Universidade Federal de Pernambuco, 50670-901, Recife, Pernambuco, Brazil}

\begin{document}
	
\maketitle

\begin{abstract}
The interaction between multilevel quantum systems and coherent radiation underlies several phenomena in modern atomic optics. The formulation and solution of the Bloch equations, which describe the dynamics of such systems, become complex as the number of levels increases. In this work, we present the Bloch Equation Generator, a free, browser-based computational tool developed to automate the generation and numerical solution of Bloch equations for systems with up to 30 levels. Users can configure the level diagram, select allowed transitions, define decay rates, and choose whether or not to apply the rotating wave approximation. The software automatically generates the complete set of equations and provides C source code for numerical solutions in both the time and frequency domains. To illustrate its applicability, we present three examples: (i) a two-level system, (ii) a $\Lambda$-type system with analysis of CPT, EIT, and the Autler–Townes effect, and (iii) a realistic 12-level system based on the Zeeman-resolved  $5S_{1/2} \to 5P_{3/2}$ transition of rubidium-87.
\end{abstract}


%
%
%
%
%

\section{\label{sec:introduction}Introduction}

The interaction between quantum systems and coherent electromagnetic fields underlies several key phenomena in modern atomic physics, such as optical resonance, electromagnetically induced transparency (EIT), coherent population trapping (CPT), Stark shift, and Autler–Townes splitting. The optical Bloch equations formalism, derived from the Liouville–von Neumann equation in the density matrix framework \cite{Cohen, Eberly}, has become one of the most widely used tools for the theoretical and computational description of such processes \cite{Gordon, Moon}.

As the number of internal levels increases --- due to the inclusion of Zeeman sublevels, multiple excitation frequencies, or decay pathways --- the formulation and numerical solution of the Bloch equations becomes very laborious, especially when done manually. For systems with $N$ levels, the number of coupled differential equations can reach $N(N+1)/2$, making the problem particularly challenging.

In this work, we present the Bloch Equation Generator (BEG), a program available on the SimuFísica platform, which automates both the generation and the numerical solution of Bloch equations for systems with up to 30 levels. The software allows users to define the diagram level, select the allowed transitions, configure decay rates, and choose whether to apply the  rotating-wave approximation. Once these parameters are set, the BEG automatically generates the full set of Bloch equations and provides C source code capable of solving them numerically, both in the time domain and as a function of field detuning.

To illustrate the capabilities of the program, we present three representative examples in this article: (i) a two-level system with a well-known analytical solution, (ii) a three-level $\Lambda$-type system in which CPT, EIT, and Autler–Townes splitting are explored, and (iii) a realistic 12-level system based on the $5S_{1/2}, F=2 \to 5P_{3/2}, F=3$ transition of $^{87}$Rb, including an analysis of $\sigma$ and $\pi$ transitions between Zeeman sublevels.

\section{\label{sec:bloch-equations}Optical Bloch equations}

We aim to describe the dynamics of an ensemble of non-interacting quantum systems with $N$ levels (e.g., atoms) subjected to coherent electromagnetic excitation. In the density matrix formalism, the time evolution of the system is governed by the Liouville–von Neumann equation \cite{sakurai}:

\begin{equation}
	\frac{\partial\hat{\rho}(t)}{\partial t} = -\frac{i}{\hslash}[\hat{H}(t), \hat{\rho}(t)],
	\label{liouville}
\end{equation}

\noindent where $\hat{\rho}(t)$ is the density matrix of the atomic system and $\hat{H}(t)$ is the Hamiltonian of the atom-field system. In the basis of eigenstates of the free atom, we write:

\begin{equation}
	\hat{H}(t) = \sum_{k=1}^N\hslash\omega_k\left| k \right\rangle \left\langle k \right|  + \hat{H}_I(t),
\end{equation}

\noindent where $\hslash\omega_k$ is the energy of the $k$-th stationary state. Assuming the electric dipole approximation, the interaction Hamiltonian takes the form:

\begin{equation}
	\hat{H}_I(t) = - \hat{p}E(t),
	\label{HI}
\end{equation}

\noindent with $\hat{p} = -e\hat{r}$ being the electric dipole operator, and $E(t)$ the total electric field, considered as a superposition of $M$ modes with frequencies $\omega_m$ and amplitudes $E_m$:

\begin{equation}
	E(t) = \sum_{m=1}^M \left( E_{m} e^{i\omega_m t} + \textrm{c.c.} \right),
\end{equation}

\noindent where c.c. denotes the complex conjugate.

Combining Eqs. (\ref{liouville})–(\ref{HI}), the time evolution of the density matrix elements $\rho_{ij}$ reads:

\begin{equation}
	\frac{\partial\rho_{ij}(t)}{\partial t} = \frac{iE(t)}{\hslash}\sum_k p_{ik}\rho_{kj} + \textrm{c.c.} \ .
	\label{liouville2}
\end{equation}

For off-diagonal elements, we introduce the following change of variables:

\begin{equation}
	\rho_{ij}(t) = \sigma_{ij}(t)e^{i\omega_m t} \ \ \ \ \ (\textrm{for} \ i \neq j),
\end{equation}

\noindent assuming that a single field mode drives each transition. Neglecting rapidly oscillating terms with frequencies $2\omega_m$ (rotating-wave approximation, RWA), we obtain the optical Bloch equations:

\begin{subequations}
	\begin{equation}
		\frac{\partial\rho_{ii}(t)}{\partial t} = i\sum_k \Omega_{ik}\sigma_{ki}(t) + \textrm{c.c.} \pm \{\textrm{decay terms}\},
	\end{equation}
	\begin{equation}
		\frac{\partial\sigma_{ij}(t)}{\partial t} = i\sum_k \Omega_{ik}\sigma_{kj}(t) + (i\delta_{ij} - \gamma_{ij})\sigma_{ij},
	\end{equation}
	\label{bloch}
\end{subequations}

\noindent where the detuning $\delta_{ij} = \omega_m - \omega_{ij}$ quantifies the deviation of the driving mode frequency from the resonance frequency of the transition $\left| i \right\rangle \rightarrow \left| j \right\rangle$, and $\Omega_{ij} = p_{ij} E_m/\hslash$, where $p_{ij} = \left\langle i | e \hat{r} | j \right\rangle$ is the dipole moment of the $\left| i \right\rangle \to \left| j \right\rangle $ transition. The ``decay terms'' are added phenomenologically to account for population relaxation, while $\gamma_{ij}$ represent coherence relaxation rates. For a three-level system with $E_i > E_j > E_l$ and allowed transitions $\left| i \right\rangle \rightarrow \left| j \right\rangle$ and $\left| j \right\rangle \rightarrow \left| l \right\rangle$, the population equations become:

\begin{subequations}
	\begin{equation}
		\frac{\partial\rho_{ii}(t)}{\partial t} = i\sum_k \Omega_{ik}\sigma_{ki}(t) - \Gamma_{ij}\rho_{ii}(t)
	\end{equation}
	\begin{equation}
		\frac{\partial\rho_{jj}(t)}{\partial t} = i\sum_k \Omega_{jk}\sigma_{kj}(t) + \Gamma_{ij}\rho_{ii}(t) - \Gamma_{jl}\rho_{jj}(t)
	\end{equation}
	\begin{equation}
		\frac{\partial\rho_{ll}(t)}{\partial t} = i\sum_k \Omega_{lk}\sigma_{kl}(t) + \Gamma_{jl}\rho_{ll}(t)
	\end{equation}
	\label{eq:decay}
\end{subequations}

\noindent where $\Gamma_{ij}$ denotes the spontaneous decay rate from $\left| i \right\rangle$ to $\left| j \right\rangle$. The relaxation rate of the coherences is given by:

\begin{equation}
    \gamma_{ij} = \frac{1}{T_{ii}} + \frac{1}{T_{jj}},
    \label{gamma}
\end{equation}

\noindent with $T_{ii} = 1/\sum_k \Gamma_{ik}$ representing the lifetime of state $\left| i \right\rangle$.

\section{\label{sec:BEG} The Bloch Equation Generator (BEG)}

\begin{figure}[ht]
    \centering
    \includegraphics[width=0.65\linewidth]{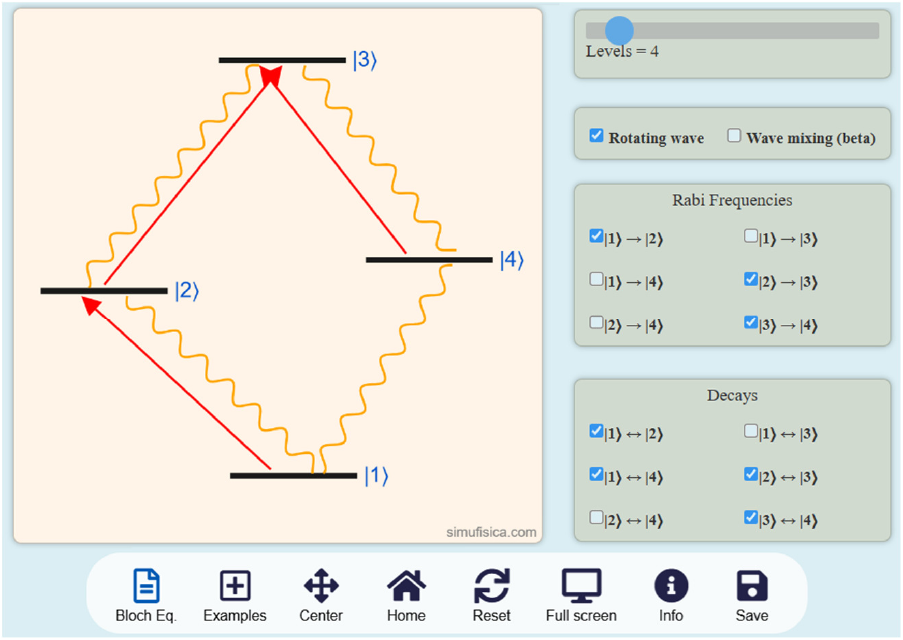}
    \caption{Bloch Equation Generator, version 1.4.2 (18/04/2025), configured for a four-level diamond-type system, with three driving field modes (red arrows) and four allowed decay pathways (yellow wavy lines). Link: \url{https://simufisica.com/en/tools/bloch-equation-generator/}.}
    \label{fig-BEG}
\end{figure}

The Bloch Equation Generator (BEG, Fig. \ref{fig-BEG}) is a web-based application developed using JavaScript, HTML, and CSS, designed to display and numerically solve the optical Bloch equations for arbitrary $N$-level quantum systems under coherent electromagnetic excitation involving multiple frequencies. The BEG is part of the SimuFísica platform\footnote{\url{https://simufisica.com/en/}}, a suite of interactive simulation tools for teaching physics at the high school and undergraduate levels in physics, engineering, and related fields \cite{Souza-2024-1-x, Souza-2024-2-x}.

Once the web page is loaded, the user selects the number of levels in the system, constrained to $2 \leq N \leq 30$. Energy level positions are adjusted interactively using a click-and-drag interface. In the \texttt{Rabi Frequencies} panel, the user defines the electric dipole-allowed transitions that can be driven by the field modes. The \texttt{Decays} panel allows specification of spontaneous relaxation pathways. The \texttt{Rotating wave} checkbox enables the application of the rotating wave approximation (RWA) to the Bloch equations.

\begin{figure}[ht]
    \centering
    \includegraphics[width=0.65\linewidth]{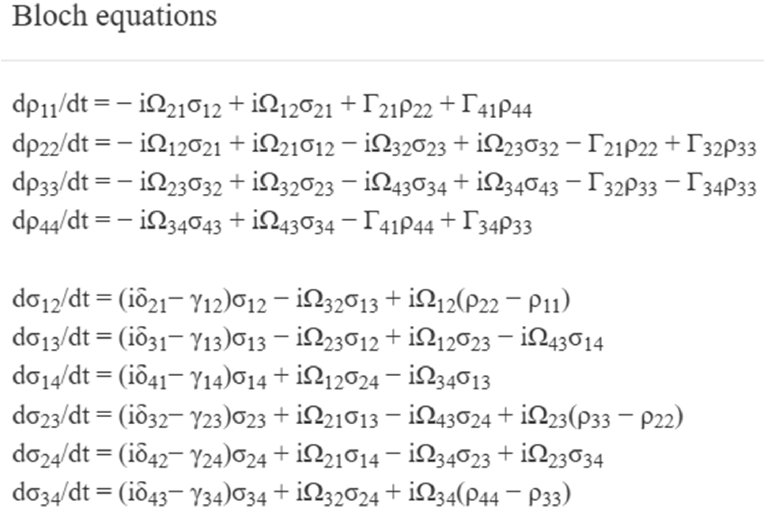}
    \caption{Bloch equations generated for the configuration shown in Fig. \ref{fig-BEG}.}
    \label{fig-bloch-equation}
\end{figure}

The Bloch equations (Fig. \ref{fig-bloch-equation}) are displayed after clicking the \texttt{Bloch Eq.} button in the toolbar. Since off-diagonal elements of the density matrix satisfy $\sigma_{ij} = \sigma^*_{ji}$, only the independent terms are shown, resulting in a total of $N(N+1)/2$ coupled differential equations.

\begin{figure}[ht]
    \centering
    \includegraphics[width=0.65\linewidth]{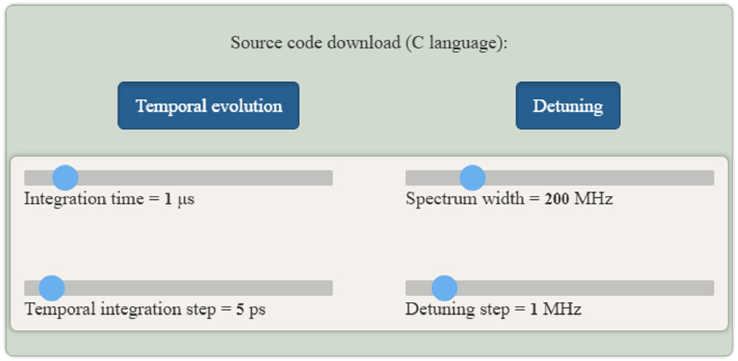}
    \caption{Code download panel in the BEG interface.}
    \label{fig-download}
\end{figure}

In the same window, users may download C source code files that numerically solve the generated Bloch equations (Fig. \ref{fig-download}). The numerical integration is implemented using the fourth-order Runge–Kutta method \cite{Press}. For time-domain solutions, the \texttt{Temporal evolution} button is used, with configuration options for the total interaction time (\texttt{Integration time}) and numerical time step (\texttt{Temporal integration step}). For spectral solutions --- where the dynamics are computed as a function of detuning of the driving modes --- the user can select the \texttt{Detuning} option, with adjustable parameters for spectral width (\texttt{Spectrum width}) and resolution (\texttt{Detuning step}).

\section{\label{sec:examples} Examples}

\subsection{Two-level system}

\begin{figure}
    \centering
    \includegraphics[width=0.60\linewidth]{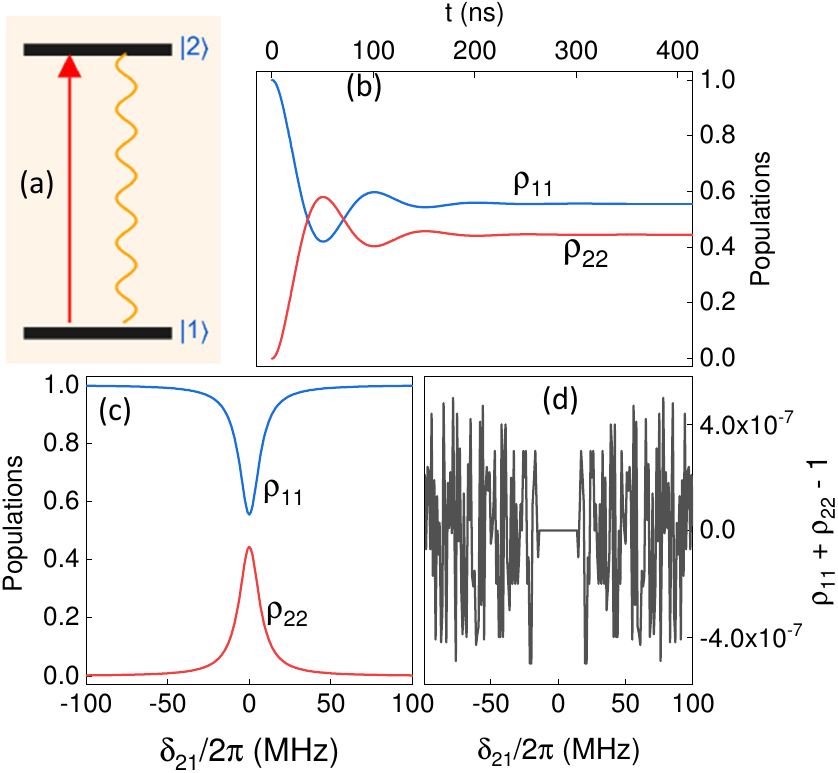}
    \caption{(a) Two-level system. (b) Time evolution of populations. (c) Steady-state populations as a function of detuning. (d) Absolute error ($\rho_{11} + \rho_{22} - 1$) as a function of $\delta_{12}$. Condition: $\Omega_{12}/\Gamma_{21} = 1$. Link to this configuration: \url{https://simufisica.com/Ndb0w}. Source code: \url{https://github.com/marcopolomoreno/bloch-equation-generator/tree/main/two-level}.}
    \label{fig-two-level}
\end{figure}

We begin by presenting results for a two-level system interacting with a cw laser beam [Fig. \ref{fig-two-level}(a)], the simplest possible system, yet rich in interesting physics, for example in the generation of nonclassically correlated photons \cite{Araujo} or even the saturated absorption spectroscopy, a rather common technique in undergraduate experimental courses \cite{Jacques}. After configuring the system in the BEG, we downloaded the source code without making any modifications, due to the simplicity of the model. $\Gamma_{21}$ was kept at 5 MHz, which is close to the 6 MHz linewidth of $5S_{1/2} \to 5P_{3/2}$ in $^{87}$Rb line \cite{steck}. We selected $\Omega_{12} = \Gamma_{21}$, corresponding to near-saturation condition, and set the detuning to zero, i.e., $\delta_{21} = 0$. The initial condition assumes full population in the ground state $\left| 1 \right\rangle$. The numerical integration parameters adopted were: total interaction time of 1~$\mu$s, time step $h = 5$~ps, and an interval of $100h$ between plotted points. These values are used throughout all examples in this work, unless stated otherwise. Listing \ref{listing-two-level} shows the relevant portion of the source code for this simulation — the links to the complete source codes are provided in the figure captions.

Figure \ref{fig-two-level}(b) shows the time evolution of the populations. The results are well-known: the populations undergo Rabi oscillations at frequency $2\Omega_{12}$ and reach the steady state at approximately $t = 2\pi/\Gamma_{21} = 200$ ns.

\begin{lstlisting}[language=C,
	caption={Source code excerpt for Fig. \ref{fig-two-level}(b) — time evolution of a two-level system.},
	label={listing-two-level},]
//Adjustments to be made (type control + F to find what needs to be adjusted)
//Adjustments 1 - Spectrum and temporal integration
//Adjustments 2 - Rabi frequencies
//Adjustments 3 - Decays
//Adjustments 4 - Initial conditions
//Adjustments 5 - Detunings
	
//Real part of Rabi frequencies
double const freqRabi = 2*Pi*5e6;

double A12 = freqRabi;

//Decay rates of excited states
double const decay = 2*Pi*5e6;

double Gamma21 = decay/1.0;

//Decay rates of coherences
double gamma12 = 0.5*decay;

//Detunings
double delta21 = 0;

//Initial populations
double pop[N+1];
pop[1] = 1.0/1.0;
pop[2] = 0;

//Initial coherences
pop[3] = 0;
pop[4] = 0;

//Integration time
double const tTotal = 1e-6;

//Time integration step
double const h = 5e-12;

//Interval between points in the graph, in units of h
int const dt = 100;
\end{lstlisting}

The population spectra as a function of detuning $\delta_{21}$, obtained in the steady-state regime, are shown in Fig. \ref{fig-two-level}(c). The parameters used were \texttt{Spectrum width = 200 MHz} and \texttt{Detuning step = 1 MHz}, as indicated in Listing \ref{listing-two-level-detuning}. The plot exhibits power broadening, with linewidths given by $\mbox{FWHM} = 2\sqrt{\gamma_{12}^2 + 4\Omega_{12}^2(\gamma_{12}/\Gamma_{21})} = 15$ MHz \cite{thesis}. A Lorentzian fit yielded $\mbox{FWHM} = 15.000004(2)$ MHz, corresponding to a relative error of $2.7 \times 10^{-5}$\%. Figure \ref{fig-two-level}(d) displays the deviation from unity of the population sum ($\rho_{11} + \rho_{22} - 1$) as a function of $\delta_{21}$, serving as a proxy for numerical accuracy. For $\Omega_{12} = \Gamma_{21}$, the maximum relative error was $5.1\times 10^{-5}$\%.

\begin{lstlisting}[language=C,
	caption={Source code excerpt for Figs. \ref{fig-two-level}(c) and (d) — population spectrum.},
	label={listing-two-level-detuning},]
//Spectrum width, in MHz
double const spectrumWidth = 200;

//Detuning step, in MHz
double const passo = 1;

delta21 = 2*Pi*passo*d*1e6;   //Field sweeping frequency
\end{lstlisting}

\subsection{Three-level $\Lambda$-type system: Coherent population trapping, electromagnetically induced transparency, and Autler-Townes splitting}

\begin{figure}
    \centering
    \includegraphics[width=0.70\linewidth]{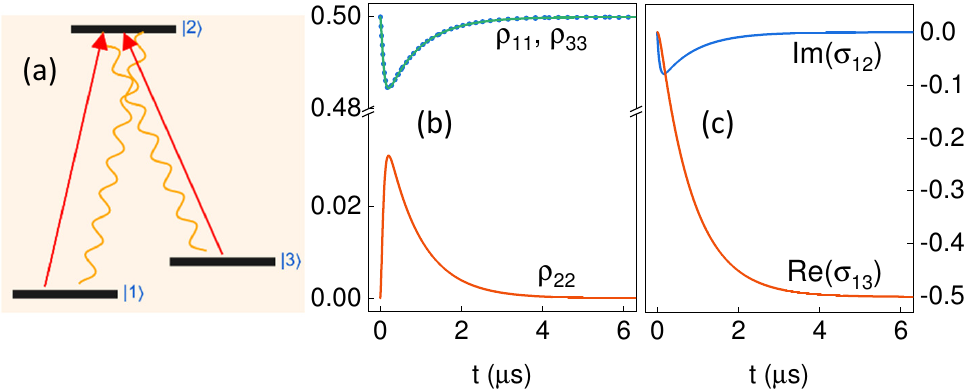}
    \caption{(a) $\Lambda$-type three-level system. (b) Time evolution of the populations. (c) Time evolution of selected coherences. Condition: $\Omega_{12} = \Omega_{32} = 0.1\times\Gamma_{21}$. Link to this configuration: \url{https://simufisica.com/DM5BP}. Source code: \url{https://github.com/marcopolomoreno/bloch-equation-generator/blob/main/three-level-\%CE\%9B/temporal-evolution.c}.}
    \label{fig-3-level-time}
\end{figure}

Our second application is based on the transition $5S_{1/2}, F=1 \to 5P_{3/2}, F=2 \to 5S_{1/2}, F=2$ in $^{87}$Rb, corresponding to the states $\left| 1 \right\rangle \to \left| 2 \right\rangle \to \left| 3 \right\rangle$ of a three-level $\Lambda$-type system [Fig. \ref{fig-3-level-time}(a)]. This is one of the most explored configurations in the literature, especially in the community of nonlinear optics. As we shall show, the same system can present numerous phenomena. The conditions chosen were $\Gamma_{21} = \Gamma_{23} = 2.5$ MHz and $\Omega_{12} = \Omega_{32} =  0.1\times\Gamma_{21}$. Both field frequencies were kept on resonance: $\delta_{21} = \delta_{32} = \delta_{31} = 0$. As the initial condition, we considered a thermal population distribution: $\rho_{11} = \rho_{33} = 0.5$ and $\rho_{22} = 0$.

It is worth noting that, once the values of $\Gamma_{ij}$ are defined, the BEG automatically determines the relaxation parameters $\gamma_{ij}$ --- according to Eq. (\ref{gamma}) --- and sets the initial populations based on the allowed transitions. A representative excerpt of the generated code is shown in Listing \ref{listing-3-level-time}. Although in this example we have defined equal decay rates from state $\left| 2 \right\rangle$ to states $\left| 1 \right\rangle$ and $\left| 3 \right\rangle$, realistic decay rates can be chosen by adjusting lines 10 and 11 of Listing \ref{listing-3-level-time}.

Figures \ref{fig-3-level-time}(b) and (c) present the numerical results for the time evolution of the populations, the real part of $\sigma_{13}$, and the imaginary part of $\sigma_{12}$. With both fields tuned to resonance, the system evolves into a dark state that decouples from the intermediate level $\left| 2 \right\rangle$, characterizing the phenomenon of coherent population trapping (CPT) \cite{arimondo}. As a result, the population remains trapped in the lower states [Fig. \ref{fig-3-level-time}(b)].

In the CPT regime, the coherence $\rho_{13}$ tends to the analytical value $\rho_{13} \to -\Omega_{12}\Omega_{23}^*/(\Omega_{12}^2 + \Omega_{23}^2)$ as $t \to \infty$. Under our simulation parameters, this leads to $\sigma_{13} \to -0.5$, as seen in the red curve in Fig. \ref{fig-3-level-time}(c). The blue curve shows the imaginary part of $\sigma_{12}$, which is directly related to the absorption of field driving the $\left| 1 \right\rangle \to \left| 2 \right\rangle$ transition.

\begin{lstlisting}[language=C,
	caption={Source code excerpt for Fig. \ref{fig-3-level-time} — time evolution of a $\Lambda$-type system.},
	label={listing-3-level-time},]
//Real part of Rabi frequencies
double const rabiFreq = 0.1*2*Pi*5e6;

double A12 = rabiFreq;
double A23 = rabiFreq;

//Decay rates of excited states
double const decay = 2*Pi*5e6;

double Gamma21 = decay/2.0;
double Gamma23 = decay/2.0;

//Decay rates of coherences
double gamma12 = 0.5*decay;
double gamma13 = 0;
double gamma23 = 0.5*decay;

//Detunings
double delta21 = 0;
double delta31 = 0;
double delta32 = 0;

//Initial populations
double pop[N+1];
pop[1] = 1.0/2.0;
pop[2] = 0;
pop[3] = 1.0/2.0;
\end{lstlisting}

\begin{figure}
    \centering
    \includegraphics[width=0.50\linewidth]{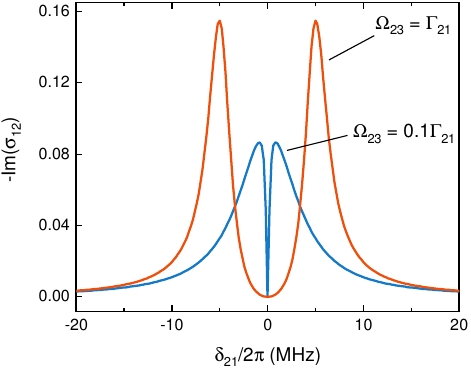}
    \caption{Absorption of the weak probe field ($\Omega_{12}/\Gamma_{21} = 0.1$) in a $\Lambda$ system as a function of $\delta_{21}$, for two values of $\Omega_{32}$. Source code: \url{https://github.com/marcopolomoreno/bloch-equation-generator/blob/main/three-level-\%CE\%9B/detuning.c}.}
	\label{fig-3-level-frequency}
\end{figure}

In Fig. \ref{fig-3-level-frequency}, we present the absorption on the $\left| 1 \right\rangle \to \left| 2 \right\rangle$ transition, with $\Omega_{12}/\Gamma_{21} = 0.1$, as a function of its detuning $\delta_{21}$, for two values of the coupling field: $\Omega_{23} = 0.1\Gamma_{21}$ (blue curve) and $\Omega_{23} = \Gamma_{21}$ (red curve). The field driving the $\left| 2 \right\rangle \to \left| 3 \right\rangle$ transition is kept on resonance ($\delta_{32} = 0$). It is important to note that, although the BEG automatically sets the two-photon conditions, it is necessary to verify whether the detunings (such as $\delta_{31} = \delta_{21} + \delta_{32}$) are correctly written, as highlighted in line 16 of Listing \ref{listing-3-level-detuning}. Furthermore, it is important to emphasize that, in systems with multiple field modes, such as in this example, the user must choose which of them should have their frequency varied, as highlighted in line 12 of Listing \ref{listing-3-level-detuning}.

The blue curve corresponds to the regime of electromagnetically induced transparency (EIT) \cite{Harris, Marangos}, while the red curve shows Autler–Townes splitting (ATS) \cite{Hao}. In the latter, the strong coupling field induces dressed states, resulting in a doublet structure in the absorption spectrum, with peaks separated by the Rabi frequency of the coupling field ($2\Omega_{23}/2\pi = 10$ MHz).

\begin{lstlisting}[language=C,
	caption={Source code excerpt for Fig. \ref{fig-3-level-frequency} — weak-field absorption in a $\Lambda$-type system.},
	label={listing-3-level-detuning},]
//Spectrum width, in MHz
double const spectrumWidth = 40;

//Detuning step, in MHz
double const passo = 0.2;

//Interaction time, in s
double const tTotal = 10e-6;

//******* Adjustments 5 - Detunings *********//
//*******************************************//
delta21 = 2*Pi*passo*d*1e6;   //Field sweeping frequency
delta32 = 0;

//Two-photon coherences
delta31 = delta21 + delta32;
//*******************************************//
\end{lstlisting}

\subsection{12-level system: $\sigma$ and $\pi$ transitions in the $5S_{1/2}, F=2 \rightarrow 5P_{3/2}, F=3$ line of $^{87}$Rb}

There are several scenarios of scientific interest in which one must consider the Zeeman atomic sublevels, e.g. when there are two driving fields with different polarizations \cite{Jadoon}. In this case, a more realistic description leads to a multi-level system, showcasing the usefulness of the BEG.

Let us consider the case of the $5S_{1/2}, F=2 \rightarrow 5P_{3/2}, F=3$ transition in $^{87}$Rb when all Zeeman sublevels are taken into account.  This renders a 12-level quantum system --- consisting of five degenerate ground states and seven degenerate excited states. This leads to a system of 78 coupled Bloch equations, whose formulation and numerical solution become laborious and error-prone without tools such as the BEG.

The allowed transitions depend on the polarization of the electromagnetic field. For circular polarization ($\sigma^\pm$), the selection rules require $\Delta m_F = \pm 1$, while for linear polarization ($\pi$), the condition is $\Delta m_F = 0$ \cite{foot}.

\begin{figure}
    \centering
    \includegraphics[width=0.65\linewidth]{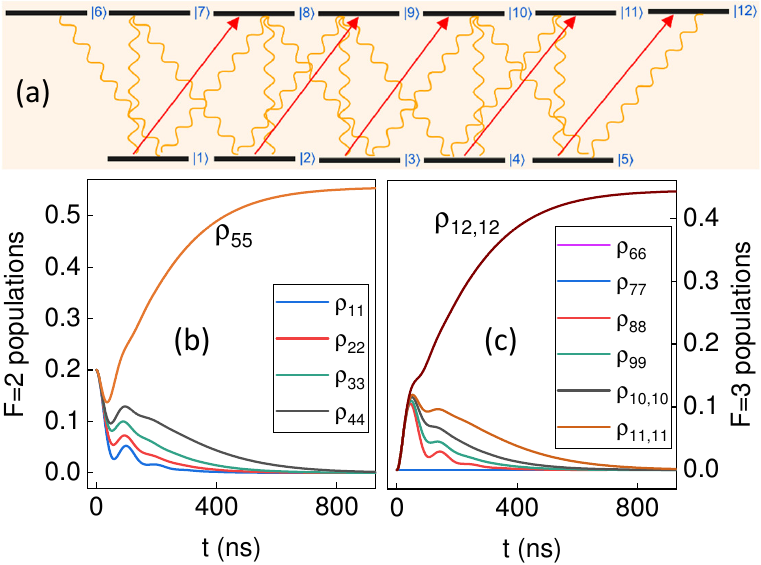}
     \caption{(a) $\sigma^+$ transitions between Zeeman sublevels in the $5S_{1/2}, F=2 \rightarrow 5P_{3/2}, F=3$ line of $^{87}$Rb. (b) Time evolution of the ground state populations. (c) Time evolution of the excited state populations. Link to this configuration: \url{https://simufisica.com/PPT3E}. Source code: \url{https://github.com/marcopolomoreno/bloch-equation-generator/blob/main/12-level/sigma/temporal-evolution.c}.}
	\label{fig5}
\end{figure}

We begin by analyzing the case where the system is excited by a circularly polarized field $\sigma^+$ [Fig. \ref{fig5}(a)], corresponding to $\Delta m_F = +1$. Allowed transitions include, for instance, $m_F = -2 \rightarrow -1$ ($\left| 1 \right\rangle \to \left| 8 \right\rangle$), $m_F = -1 \rightarrow 0$ ($\left| 2 \right\rangle \to \left| 9 \right\rangle$), and so on. For spontaneous decay, we consider $\Delta m_F = 0, \pm 1$.

Figures \ref{fig5}(b) and \ref{fig5}(c) show the time evolution of populations in the ground and excited states, respectively. For simplicity, we assumed equal lifetimes for all excited states ($\sum_i \Gamma_{ij} \equiv \Gamma$) and equal Rabi frequencies for all allowed transitions: $\Omega_{ij} \equiv \Omega = \Gamma$. The field was kept on resonance: $\delta_{ji} \equiv \delta = 0$. Listing \ref{listing-12-level-time} shows representative decay parameters, coherence rates, and initial populations configured automatically by the BEG.

Under these conditions, we observe that the population becomes concentrated in the Zeeman sublevels with the highest magnetic quantum number $m_F$ ($\left| 5 \right\rangle$ and $\left| 12 \right\rangle$), effectively reducing the system to a two-level configuration \cite{Lezama}. This behavior is typical of $\sigma$ transitions with $\Delta F = +1$, especially when $\Omega > \Gamma/2$.

\begin{figure}
    \centering
    \includegraphics[width=0.80\linewidth]{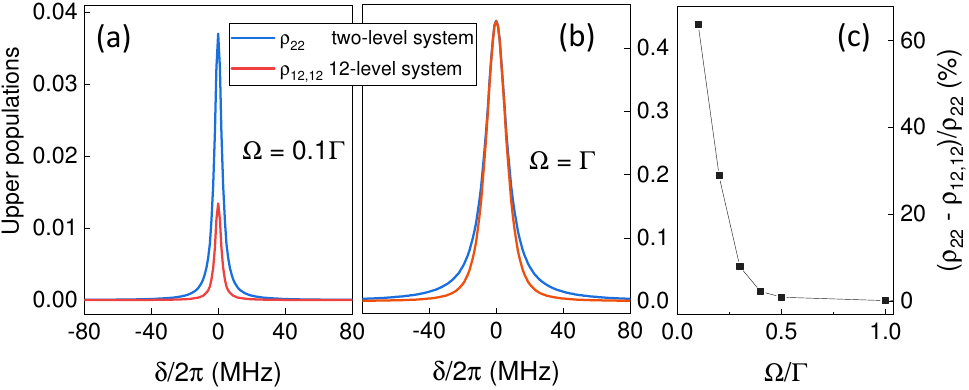}
    \caption{Comparison between the excited-state population in a two-level system ($\rho_{22}$) and in the 12-level system with $\sigma^+$ transitions ($\rho_{12,12}$). (a) $\Omega = 0.1\Gamma$. (b) $\Omega = \Gamma$. (c) Percentage difference, $(\rho_{22} - \rho_{12,12})/\rho_{22}$, as a function of $\Omega/\Gamma$ at $\delta = 0$. Source code: \url{https://github.com/marcopolomoreno/bloch-equation-generator/blob/main/12-level/sigma/detuning.c}.}
    \label{fig-2-vs-12-level}
\end{figure}

This equivalence is illustrated in Fig. \ref{fig-2-vs-12-level}, which compares the excited-state spectrum of a two-level system ($\rho_{22}$) with the population of the last excited Zeeman sublevel ($\rho_{12,12}$) in the 12-level system. For $\Omega/\Gamma = 0.1$ [Fig. \ref{fig-2-vs-12-level}(a)], the difference is significant. For $\Omega = \Gamma$ [Fig. \ref{fig-2-vs-12-level}(b)], the curves become nearly identical. Figure \ref{fig-2-vs-12-level}(c) shows the relative percentage difference between the populations at $\delta = 0$ as a function of $\Omega/\Gamma$, which approaches zero for values above 0.5.

\begin{lstlisting}[language=C,
	caption={Source code excerpt for Figs. \ref{fig5}(b) and (c).},
	label={listing-12-level-time},]
//Decay rates of excited states
double const decay = 2*Pi*5e6;

double Gamma61 = decay/1.0;
double Gamma71 = decay/2.0;
double Gamma81 = decay/3.0;
double Gamma72 = decay/2.0;
//...
//Decay rates of coherences
double gamma12 = 0;
double gamma13 = 0;
double gamma14 = 0;
double gamma15 = 0;
double gamma16 = 0.5*decay;
double gamma17 = 0.5*decay;
double gamma18 = 0.5*decay;
//...
//Initial populations
double pop[N+1];
pop[1] = 1.0/5.0;
pop[2] = 1.0/5.0;
pop[3] = 1.0/5.0;
pop[4] = 1.0/5.0;
pop[5] = 1.0/5.0;
\end{lstlisting}

\begin{figure}[ht]
    \centering
    \includegraphics[width=0.65\linewidth]{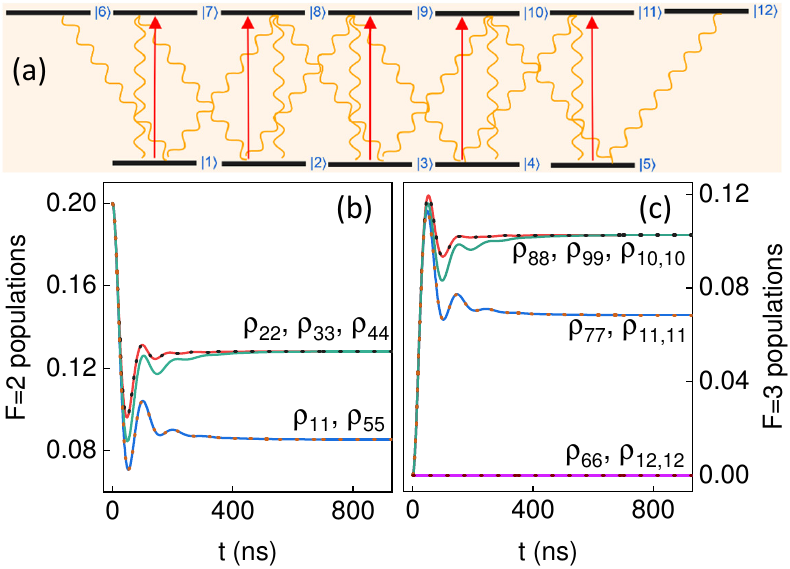}
    \caption{(a) $\pi$ transitions between Zeeman sublevels in the $5S_{1/2}, F=2 \rightarrow 5P_{3/2}, F=3$ line of $^{87}$Rb. (b) Time evolution of the ground-state Zeeman populations. (c) Time evolution of the excited-state Zeeman populations. Link to this configuration: \url{https://simufisica.com/nwOow}. Source code: \url{https://github.com/marcopolomoreno/bloch-equation-generator/blob/main/12-level/pi/temporal-evolution.c}.}
    \label{fig-12-level-pi}
\end{figure}

Finally, we consider the case of $\pi$ transitions [Fig. \ref{fig-12-level-pi}(a)], where $\Delta m_F = 0$. In this configuration, the population is not funneled into just two Zeeman sublevels. Instead, atoms distribute among all magnetic sublevels, except for the upper levels where $m_F = \pm F$, as shown in Figs. \ref{fig-12-level-pi}(b) and (c). For $\Omega \geq \Gamma$, the excited-state populations tend to become evenly distributed among $m_F = 0, \pm 1, \ldots, \pm (F-2)$, with lower occupancy in the extreme sublevels $m_F = \pm (F-1)$ \cite{Zigdon}.

\section{\label{sec:conclusions} Conclusions}

We have presented the Bloch Equation Generator, a computational tool developed to automate the generation and numerical solution of the Bloch equations for arbitrary quantum systems with up to 30 levels. The software allows users to configure the system's structure, the allowed transitions and decays, as well as the numerical integration parameters. Based on these inputs, the BEG displays the corresponding Bloch equations and generates C source code for both time-domain and frequency-domain solutions. The number of equations grows rapidly with the number of levels [$N(N+1)/2$], highlighting the importance of the automation provided by the tool.

The examples discussed --- from the two-level system to the realistic 12-level $^{87}$Rb system --- demonstrate how the BEG can be used to analyze quantum systems with a wide range of configurations. We hope the software proves useful for research involving the solution of Bloch equations in complex multilevel systems.

Future developments aim to further enhance the usability and performance of the tool. One planned feature is the ability to display steady-state spectral plots directly within the browser interface, eliminating the need to download and compile C source code for visualization. Additionally, support for CUDA C is being considered, enabling GPU acceleration on Nvidia hardware \cite{Demeter} to significantly reduce computation time in frequency-domain solutions.

\section*{Acknowledgements}

This research was funded by Coordena\c{c}\~{a}o de Aperfei\c{c}oamento de Pessoal de N\'{\i}vel Superior (CAPES - PROEX Grant 534/2018, No. 23038.003382/2018-39), 
Fundação de Amparo ao Desenvolvimento das Ações Científicas e Tecnológicas e à Pesquisa do Estado de Rondônia (FAPERO, Grant 36214.577.20546.20102023), Funda\c{c}\~{a}o de Amparo \`{a} Pesquisa do Estado de S\~{a}o Paulo (FAPESP - Grant 2021/06535-0), and Universidade Federal de Rondônia (UNIR, Grant 23118.006316/2024-79). M. P. M. de Souza acknowledges financial support from CNPq (Grant 304017/2022-1). A. A. C. de Almeida acknowledges financial support from CNPq (Grant 351985/2023-9) and Office of Naval Research (ONR - Grant N62909-23-1-2014). S. S. Vianna acknowledges financial support from CNPq (Grant 307722/2023-6).

\bibliographystyle{iopart-num}
\bibliography{biblio}

\end{document}